\documentclass[preprint2,numberedappendix]{emulateapj-rtx4}

\usepackage{graphicx}
\usepackage{natbib}

\usepackage{bm,times,color,slashbox}
\graphicspath{{./fig/}{./png/}}

\newcommand{\EQ}{\begin{equation}}
\newcommand{\EN}{\end{equation}}
\newcommand{\EQA}{\begin{eqnarray}}
\newcommand{\ENA}{\end{eqnarray}}

\newcommand{\Eq}[1]{Equation~(\ref{#1})}
\newcommand{\Eqs}[2]{Equations~(\ref{#1}) and~(\ref{#2})}

\newcommand{\Sec}[1]{Section~\ref{#1}}

\newcommand{\Fig}[1]{Figure~\ref{#1}}

\newcommand{\Figp}[2]{Figure~\ref{#1}(#2)}

\newcommand{\Tab}[1]{Table~\ref{#1}}
\newcommand{\Tabs}[2]{Tables~\ref{#1} and \ref{#2}}

\newcommand{\bra}[1]{\langle #1\rangle}

{}
{}
{}

{}
{}
{}
{}
{}
{}
{}
{}
{}
{}
{}
{}
{}
{}
{}
{}
{}
{}
{}
{}
{}

{}
{}
{}

{}

{}
{}

\newcommand{\pphi}{\hat{\bm{\phi}}}

\newcommand{\rr}{\mbox{\boldmath $r$} {}}

\newcommand{\bb}{\bm{b}}

\newcommand{\jj}{\mbox{\boldmath $j$} {}}

\newcommand{\FF}{\mbox{\boldmath $F$} {}}

\newcommand{\GG}{\mbox{\boldmath $G$} {}}

\newcommand{\nab}{\mbox{\boldmath $\nabla$} {}}

\newcommand{\ii}{{\rm i}}

\newcommand{\dd}{{\rm d} {}}

\def\degr{\hbox{$^\circ$}}

\def\Rey{\mbox{\rm Re}}

\def\hrms{h_{\rm rms}}

\newcommand{\G}{\,{\rm G}}

\newcommand{\kG}{\,{\rm kG}}

\newcommand{\nm}{\,{\rm nm}}
\newcommand{\cm}{\,{\rm cm}}

\newcommand{\km}{\,{\rm km}}

\newcommand{\Mm}{\,{\rm Mm}}

\newcommand{\yjmp}[3]{ #1, {JMP,} {#2}, #3}
\newcommand{\yapj}[3]{ #1, {ApJ,} {#2}, #3}

\newcommand{\yapjl}[3]{ #1, {ApJ,} {#2}, #3}

\newcommand{\yan}[3]{ #1, {Astron.\ Nachr.,} {#2}, #3}

\newcommand{\yana}[3]{ #1, {A\&A,} {#2}, #3}
\newcommand{\dana}[3]{ #1, {A\&A}, DOI:#2, arXiv:#3}

\newcommand{\yjetp}[3]{ #1, {Sov.\ Phys.\ JETP,} {#2}, #3}

\newcommand{\yaraa}[3]{ #1, {ARA\&A,} {#2}, #3}

\newcommand{\yprl}[3]{ #1, {Phys.\ Rev.\ Lett.,} {#2}, #3}

\newcommand{\ymn}[3]{ #1, {MNRAS,} {#2}, #3}

\newcommand{\ysph}[3]{ #1, {Solar Phys.,} {#2}, #3}

\newcommand{\yprd}[3]{ #1, {Phys.\ Rev.\ D,} {#2}, #3}

\newcommand{\yjour}[4]{ #1, {#2}, {#3}, #4}

\newcommand{\ybook}[3]{ #1, {#2} (#3)}
\newcommand{\yproc}[5]{ #1, in {#3}, ed.\ #4 (#5), #2}

\newcommand{\arxiv}[2]{ #1, arXiv:#2}

\begin{document}

\title{A global two-scale helicity proxy from $\pi$-ambiguous solar magnetic fields}

\author{
Axel Brandenburg$^{1,2,3,4}$\thanks{E-mail:brandenb@nordita.org}
}

\affil{
$^1$Nordita, KTH Royal Institute of Technology and Stockholm University, Roslagstullsbacken 23, SE-10691 Stockholm, Sweden\\
$^2$Department of Astronomy, AlbaNova University Center, Stockholm University, SE-10691 Stockholm, Sweden\\
$^3$JILA and Laboratory for Atmospheric and Space Physics, University of Colorado, Boulder, CO 80303, USA\\
$^4$McWilliams Center for Cosmology \& Department of Physics, Carnegie Mellon University, Pittsburgh, PA 15213, USA
}

\date{~$Revision: 1.123 $~~}

\begin{abstract}
If the $\alpha$ effect plays a role in the generation of the Sun's
magnetic field, the field should show evidence of magnetic helicity of
opposite signs at large and small length scales.
Measuring this faces two challenges: (i) in weak-field
regions, horizontal field measurements are unreliable because of the
$\pi$ ambiguity, and (ii) one needs a truly global approach to computing
helicity spectra in the case where one expects a sign reversal across
the equator at all wavenumbers.
Here we develop such a method using spin-2 spherical harmonics to
decompose the linear polarization in terms of the parity-even
and parity-odd $E$ and $B$ polarizations, respectively.
Using simple one- and two-dimensional models, we show that the product
of the spectral decompositions of $E$ and $B$, taken at spherical
harmonic degrees that are shifted by one, can act as a proxy of the
global magnetic helicity with a sign that represents that in the
northern hemisphere.
We then apply this method to the analysis of solar synoptic vector
magnetograms, from which we extract a pseudo-polarization corresponding
to a ``$\pi$-ambiguated'' magnetic field, i.e., a magnetic field vector
that has no arrow.
We find a negative sign of the global $EB$ helicity proxy at spherical
harmonic degrees of around 6.
This could indicate a positive magnetic helicity at large length scales,
but the spectrum fails to capture clear evidence of the well-known
negative magnetic helicity at smaller scales.
This method might also be applicable to stellar and Galactic polarization data.
\end{abstract}

\keywords{
magnetic fields --- polarization --- techniques: polarimetric --- Sun: dynamo
}

\section{Introduction}

The magnetic field of the Sun and other late-type stars is known to
have, on average, opposite signs of magnetic helicity in the northern
and southern hemispheres \citep{See90,PCM95}.
There is also the possibility of the field being bihelical \citep{BB03}
with a sign change of the magnetic helicity at large length scales.
To detect this in the Sun, one would need to measure {\em spectra} of
magnetic helicity, but this is made complicated by the fact that the
solar surface also displays a systematic north--south variation with
opposite signs in the two hemispheres.
To capture this correctly, a global approach must be adopted that takes
the systematic north--south variation into account.
This is done by utilizing what is known as a two-scale approach
\citep{RS75}.
Here, one scale is that of the large-scale hemispheric modulation, and
the other is the scale of the turbulence, which in itself comprises an
entire range of length scales.
In that approach, one can compute a spectrum covering both north and
south, while taking a systematic north--south variation into account
as if both hemispheres looked just like the northern hemisphere
\citep[][hereafter BPS]{BPS17}.

The problem with the standard two-scale approach is that it is only
a {\em semi-global} one.
Technically, it is still Cartesian in that the solar surface magnetic
field is represented in the Lambert cylindrical equal-area projection.
In a proper global approach, by contrast, one would need to employ a
spherical harmonics decomposition, but this must be done in such a way
that the systematic north--south variation can still be taken into account.

In this paper, a simple heuristic modification to the usual spherical
harmonics spectra is being proposed.
It is based on the idea that in the semi-global two-scale approach, the
helicity spectrum is computed as the product of the magnetic field and
its vector potential at wavenumbers that are offset for the two fields
by a small amount that corresponds to the wavenumber of the large-scale
hemispheric modulation.
Analogously, for spherical harmonics spectra, one should consider the
product of the two terms at spherical harmonic degrees that are shifted
by one.
This idea is then adapted to analyzing also the parity-even and parity-odd
contributions to the linear polarization \citep{Kamion97,SZ97}.
The reason for using such a decomposition is that there are large
uncertainties owing to the $\pi$ ambiguity of the magnetic field in
weak-field regions of the Sun.
This ambiguity reflects the fact that polarization ``vectors'' have
neither head nor tail.

Various disambiguation procedures are available \citep{Sak85,Geo05,
Hoe14,RA14}, but they tend to fail in regions far away from sunspots,
where the magnetic field is weak.
To avoid any bias, the random disambiguation method is often employed
\citep{Liu17}.
This is justified when the Stokes $Q$ and $U$ parameters are dominated
by noise but, if this were indeed the case, it should not be possible
to detect any systematic north--south dependence of the parity-odd $EB$
correlation from weak-field regions.
It is also clear that any magnetic helicity derived from a randomly
disambiguated magnetic field may itself be random and would therefore
be unreliable.

The proper way out of this problem of obtaining a qualitative measure
of the Sun's magnetic helicity from $\pi$-ambiguous magnetic fields is
to work directly with the original linear polarization.
This has already been attempted by determining the rotationally invariant
parity-even and parity-odd contributions, or $E$ and $B$ polarizations,
respectively, from the Stokes $Q$ and $U$ parameters
\citep[][hereafter BBKMRPS]{BBKMR19}.
This decomposition yields a field that is parity even, i.e., statistically
mirror symmetric, and another one that is parity odd, i.e., statistically
mirror antisymmetric \citep{Kamion97,SZ97}.
The relevant diagnostic quantity is usually the cross-correlation
of the spectral representations of $E$ and $B$ \citep{KR05,KMLK14,Bracco19}.

Attempts to analyze solar $E$ and $B$ polarizations have not yet produced
a nonvanishing cross-correlation (BBKMRPS).
However, this could be caused by their method still being provisional
in that only a semi-global approach was used to deal with the fact that
the sign of the cross-correlation is systematically different in the
northern and southern hemispheres.
It was always clear that a proper analysis should involve a decomposition
into spherical harmonics.
More precisely, the linear polarization parameters $Q$ and $U$ must
be decomposed into what is known as spin-2 spherical harmonics, which
have the appropriate transformation properties for linear polarization
\citep{Kamion97,SZ97}; see \cite{Dur08} for a textbook on the subject.
While this method is now routinely applied in cosmology using data from
the {\em Planck} satellite \citep{Akrami18}, it has not yet been adapted to
the case where one expects there to be a global sign change
of magnetic helicity about the equator.
In that case, we employ the spherical harmonics decomposition of
$E$ and $B$, which yields $\tilde{E}_{\ell m}$ and $\tilde{B}_{\ell m}$,
respectively.
We then compute their product at spherical harmonic
degrees that are shifted by one, i.e., we compute
$\tilde{E}_{\ell m}\tilde{B}_{\ell+1\, m}^\ast$.
We also compute $\tilde{E}_{\ell m}\tilde{B}_{\ell-1\, m}^\ast$, which
we shall show to be a better proxy of the expected magnetic helicity
spectrum than the former one.

The work of BBKMRPS suffered from another problem in that the
publicly available polarization data were not cleaned and corrected
to the same extent as those finally used to compute the Sun's magnetic
field \citep{Hughes}.
For example, the quality of the images varied across the solar disk.
Furthermore, proper line fits to solar atmosphere models have not been
performed.
Therefore, there is a possibility of small shifts in frequency that
could affect the resulting $Q$ and $U$ signals.
In particular, the magnetic field can have different strengths at
different geometrical depths, giving rise to more complicated spectral
line profiles that are usually fully accounted for in the inversion
pipelines \citep{Hoe14}, but they were ignored in the more rudimentary
analysis of BBKMRPS.
A legitimate way out of this additional problem is to use the full
solar magnetic field inversion along with its questionable disambiguated
magnetic field and make it ambiguous again!
We can do this by computing a synthetic (or pseudo) linear polarization
from the horizontal magnetic field.
Such work is already in progress (A.\ Prabhu, in preparation), but it
is still local and constrained to finite patches in one hemisphere,
as was done in the works of BPS and \cite{Singh18}.
Here, by contrast, we employ a novel analysis using spin-2 spherical
harmonics to compute a global cross-correlation spectrum.

We begin by testing the global two-scale approach and its ability to
extract a unique spectrum by using data from both hemispheres at the
same time.
In \Sec{Axisymmetric}, we first construct simple axisymmetric fields to
study the effects of a global sign change of the magnetic helicity.
In \Sec{NonAxisymmetric}, we consider nonaxisymmetric magnetic fields
to verify the numerical approach.
In \Sec{Applications}, we use synoptic magnetograms from Carrington
rotations (CRs) 2161 to 2163, for which a semi-global helicity spectrum
was previously determined (BPS).
We discuss the relevance of our results for dynamo theory in
\Sec{Implications} and conclude with the broader implications of the
present work in \Sec{Concl}.

\section{An axisymmetric example}
\label{Axisymmetric}

\subsection{Representation of the magnetic field}
\label{RepresentationAxisymmetric}

It is useful to begin with a simple example that is similar in spirit
to the one-dimensional example used in BPS (see their Figure~1), where
the magnetic helicity density shows a sign change in the middle of
the domain.
For this purpose, we restrict ourselves to an axisymmetric magnetic field,
which can be written in the form
\EQ
\bb=\nab\times(a_\phi\pphi)+b_\phi\pphi,
\label{axisymb}
\EN
where $r$ and $\theta$ are radius and colatitude, $a_\phi(r,\theta)$
is the toroidal component of the magnetic vector potential, and
$b_\phi(r,\theta)$ is the toroidal component of the magnetic field itself.
The proper expansion of $a_\phi$ and $b_\phi$ is in terms of the
associated Legendre polynomials $P_l^1(\cos\theta)$ as
\EQ
a_\phi\pphi=\sum_{\ell=1}^{N_\ell}\tilde{a}_\ell(r)P_\ell^1(\cos\theta),\quad
b_\phi\pphi=\sum_{\ell=1}^{N_\ell}\tilde{b}_\ell   P_\ell^1(\cos\theta),
\label{axisymbExpand}
\EN
where $N_\ell$ determines the truncation level.
The two horizontal magnetic field components on the surface of the
sphere at $r=R$, say, are then given by
\EQ
b_\theta(\theta)=-{1\over R}\sum_{\ell=1}^{N_\ell}
\frac{\partial}{\partial r}(r\tilde{a}_\ell)P_\ell^1(\cos\theta),
\label{bt_rder}
\EN
\EQ
b_\phi(\theta)=\sum_{\ell=1}^{N_\ell}
\tilde{b}_\ell P_\ell^1(\cos\theta).
\EN
Even if $\tilde{a}_\ell(r)$ were independent of $r$, the values of
$b_\theta$ would be finite because of the $r$ factor under the derivative.
At the surface, however, it is more likely that
$\tilde{a}_\ell(r)$ decays with $r$ as a power law, for example like
$r^{-(\ell+1)}$, as it would if the exterior magnetic field was a
potential field \citep{KR80}.
In such a case, $b_\phi$ would normally vanish, but this will not be
assumed here, because then the magnetic field would have vanishing
helicity.
Specifically, we are interested in a field with globally antisymmetric
magnetic helicity, so we assume that $b_\phi$ remains finite at $r=R$.

\subsection{Opposite helicities in the two hemispheres}
\label{Opposite}

In BPS, we constructed a magnetic field with globally antisymmetric
helicity by having the two horizontal field components with a relative
wavenumber shift that corresponds to the scale of the latitudinal
variation of the magnetic helicity.
This corresponds to the two components having an $\ell$ value that is
different by one.
In the present case, we choose $b_\ell=b_0$ and $a_\ell=-b_0 R/\ell$,
with some general amplitude factor $b_0$, so
\EQ
b_\theta(\theta)=-b_0 P_\ell^1(\cos\theta),\quad
b_\phi(\theta)=b_0 P_{\ell+1}^1(\cos\theta).
\label{HelField}
\EN
Analogously to BBKMRPS, we compute the complex linear polarization
at $r=R$ as
\EQ
p\equiv Q+\ii U=-\epsilon\,(b_\theta+\ii b_\phi)^2,
\label{Emissivity}
\EN
where $\epsilon$ is the emissivity, which is here assumed to be constant.
The minus sign in front of $\epsilon$ accounts for the fact that
polarization is related to the electric field, which is at right angles
to the magnetic field.

\subsection{Spin-weighted spherical harmonics}

Next, we decompose $p(\theta)$ into spin-weighted spherical harmonics
\citep{Kamion97,SZ97}.
The following expressions readily apply to the nonaxisymmetric case
where the complex polarization also depends on longitude $\phi$,
i.e., $p=p(\theta,\phi)$.
The spin-weighted spherical harmonics are computed as \citep{Goldberg67}
\EQ
_s Y_{\ell m}(\theta,\phi)=_s\!{\cal N}_{\ell m}
\, _s{\cal P}_{\ell m}\big(\!\sin(\theta/2),\cos(\theta/2)\big)
\,e^{\ii m\phi},
\EN
where
\EQ
_s {\cal N}_{\ell m}=(-1)^m
\sqrt{{2\ell+1\over4\pi}\,
{(\ell+m)!\over(\ell+s)!}{(\ell-m)!\over(\ell-s)!}}
\EN
is a normalization factor,
\EQ
_s{\cal P}_{\ell m}(x,y)=x^{2\ell}\,\sum_{r=0}^{\ell-s}
{_{rs}}{\cal M}_{\ell m} \, (y/x)^{2r+s-m}
\label{probl}
\EN
are polynomials of $x$ and $y/x$, and
\EQ
_{rs}{\cal M}_{\ell m}=
\pmatrix{\ell-s\cr r} \pmatrix{\ell+s\cr r+s-m}
(-1)^{\ell-r-s}
\EN
is yet another normalization factor, where the binomials are defined to
be zero when either of the arguments or their difference is nonpositive.
In \Tab{Tab1}, we list a few selected spin-2 spherical harmonics.

\begin{table}[t!] \caption{
The first few spin-2 spherical harmonics.
}\centerline{\begin{tabular}{lcc}
\hline
\hline
$\ell$ & $m$ & $_2 Y_{\ell m}(\theta,\phi)$ \\
\hline
2 &   0    & $(3/4)\sqrt{5/6\pi} \sin^2\theta$ \\
2 & $\pm1$ & $-(1/4)\sqrt{5/\pi} \sin\theta(1\mp\cos\theta)e^{\pm\ii\phi}$ \\
2 & $\pm2$ & $(1/8)\sqrt{5/\pi} (1\mp\cos\theta)^2e^{\pm2\ii\phi}$ \\
3 &   0    & $(1/4)\sqrt{105/2\pi} \sin^2\theta\cos\theta$ \\
4 &   0    & $(15/4)\sqrt{9/10\pi} \sin^2\theta[1-(7/6)\sin^2\theta]$ \\
4 & $\pm3$ & $(1/4)\sqrt{63/2\pi} \sin\theta(1\mp\cos\theta)
[(1\mp\cos\theta)/2-\sin^2\theta]e^{\pm3\ii\phi}$ \\
\label{Tab1}\end{tabular}}\end{table}

The numerical application of \Eq{probl} can become problematic
in the first (second) quadrant for $m>0$ ($m<0$) and $\theta\to0$
($\theta\to\pi$), because the sum has large terms of alternating sign.
This is not the case in the correspondingly other quadrant.
However, for the cases listed in \Tab{Tab1}, we observe that
$_2Y_{\ell m}(\theta,\phi)={_2Y}_{\ell\, -m}(\pi-\theta,\phi)$,
although this relation is not generally true.

\subsection{Spin-2 spherical harmonics decomposition}
\label{Spin2}

We now compute the spin-2 spherical harmonics representation of
$E+\ii B$ in terms of $Q+\ii U$ as \citep{Kamion97,SZ97,Dur08,KK16}
\EQ
\tilde{R}_{\ell m}=\int_{4\pi}
(Q+\ii U)\,_2 Y_{\ell m}^\ast(\theta,\phi)\,
\sin\theta\,\dd\theta\,\dd\phi,
\label{EBfromQU}
\EN
and define $\tilde{E}_{\ell m}=(\tilde{R}_{\ell m}+\tilde{R}_{\ell,\,-m}^\ast)/2$
as the parity-even part and
$\tilde{B}_{\ell m}=(\tilde{R}_{\ell m}-\tilde{R}_{\ell,\,-m}^\ast)/2\ii$
as the parity-odd part in spectral space, where the asterisk means complex
conjugation, and commas have been used to separate $\ell$ from $-m$.
In the axisymmetric case, we have $m=0$ and drop the index $m$.
Furthermore, $\tilde{E}_{\ell}$ and $\tilde{B}_{\ell}$ are then real.
It should also be noted that our coefficients $\tilde{E}_{\ell m}$ and $\tilde{B}_{\ell m}$ are sometimes defined with the opposite sign
\citep[see, e.g.][]{ZS97}.
Here we follow the sign convention of the textbook by \cite{Dur08}.

\begin{figure*}[t!]\begin{center}
\includegraphics[width=\textwidth]{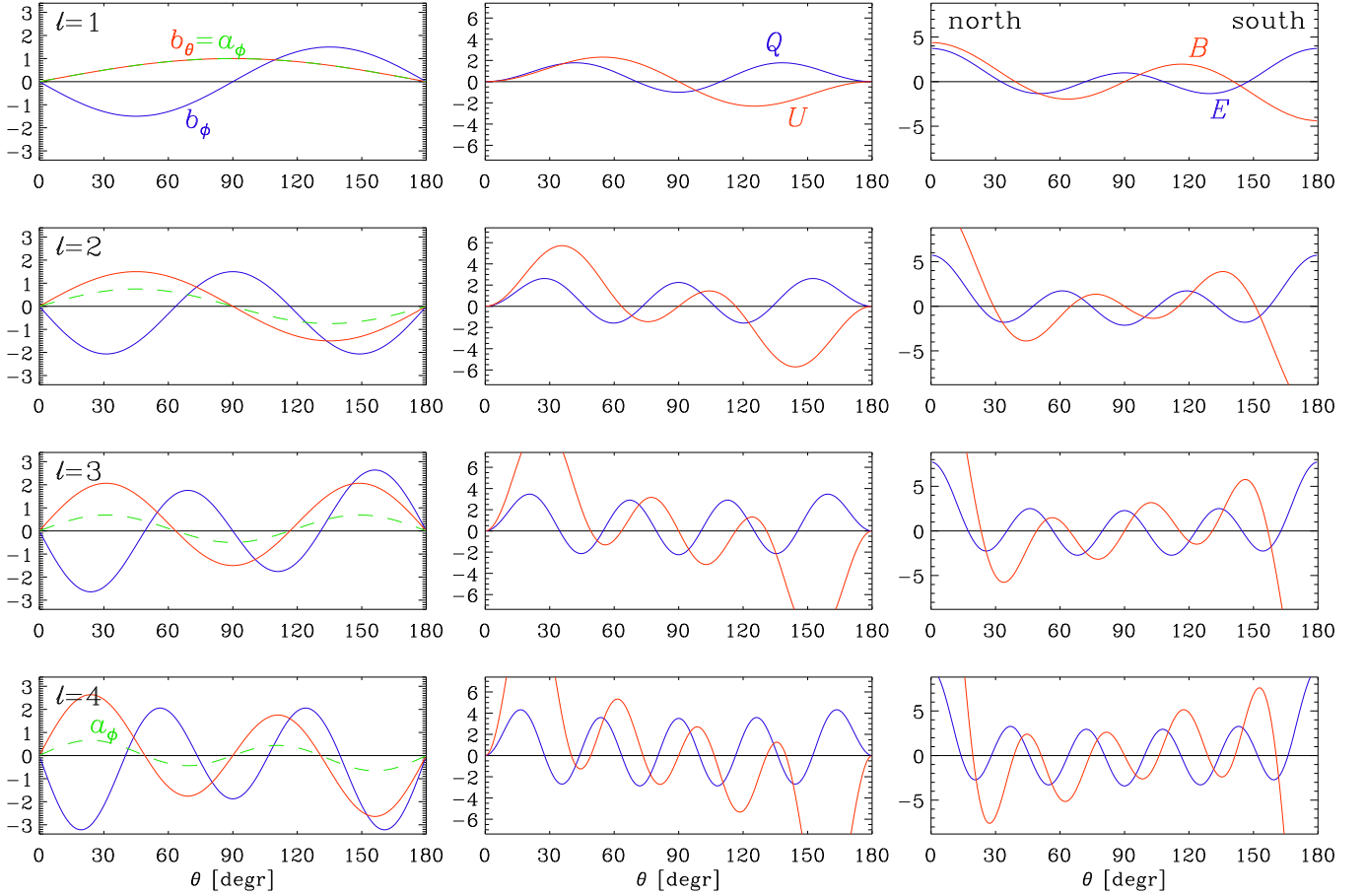}
\end{center}\caption[]{
Latitudinal dependence of $a_\phi$ (dashed green), $b_\phi$ (blue), and
$b_\theta$ (red) (left column), $Q$ (blue) and $U$ (red) (middle column),
and $E$ (blue) and $B$ (red) (right column) for the one-dimensional model.
}\label{p2_4panels}\end{figure*}

The spatial dependencies of $E(\theta,\phi)$ and $B(\theta,\phi)$ are given by
the real and imaginary parts of the inverse transform, $R$, i.e.,
\EQ
E+\ii B\equiv R=\sum_{\ell=2}^{N_\ell}\sum_{m=-\ell}^{\ell}
\tilde{R}_{\ell m} Y_{\ell m}(\theta,\phi).
\EN
It turns out that for a magnetic field given by \Eq{HelField}, finite
values of $\tilde{E}_\ell$ are only obtained for even $\ell$ ($\ell\ge2$),
while finite values of $\tilde{B}_\ell$ are only obtained for odd $\ell$
($\ell\ge3$).
In \Fig{p2_4panels}, we show the $\theta$ dependence of the components
of the two surface components of $\bb$, as well as the fields $(Q,U)$
and $(E,B)$ for several values of $\ell$.

In \Fig{p2_4panels}, we also show $a_\phi$, which is just
$b_\theta R/\ell$, where the $\ell$ factor comes from the $r$ derivative
in \Eq{bt_rder} and the fact that $\tilde{a}_\ell(r)\propto r^{-(\ell+1)}$.
We choose $\tilde{b}_\ell=-\ell\tilde{a}_\ell/R=b_0$.
In that case, positive contributions to the local magnetic helicity
density, $h(\theta)=2a_\phi b_\phi$ \citep{BDS02}, come from
$\pi/2\leq\theta\leq\pi$, i.e., from the southern hemisphere.
Negative contributions come from the northern hemisphere.
This corresponds to what is seen on the Sun for the small-scale field,
i.e., the field with $k>0.1\Mm^{-1}$.
We emphasize here that the corresponding scale,
$2\pi/0.1\Mm^{-1}\approx60\Mm$, is obviously not small by some
standards, but it is small relative to the large-scale field of the
Sun that manifests itself through the 11 yr cycle and the hemispheric
antisymmetry of the mean toroidal field.

To distinguish the spherical harmonic degrees of the magnetic field from
those of the $E$ and $B$ polarization, we denote the former with a prime
as $\ell'$.
In order to have negative (positive) contributions to the local magnetic
helicity density in the northern (southern) hemisphere, we now choose
analogously to \Eq{HelField},
\EQ
\tilde{a}_{\ell}=-\delta_{\ell\,\ell'},\quad
\tilde{b}_{\ell}=\delta_{\ell\,\ell'+1},
\label{aellbell}
\EN
for selected values of $\ell'$.
Thus, for $\ell'=1$, for example, we have $\tilde{a}_1=-1$ and
$\tilde{b}_2=1$ as the only two nonvanishing coefficients, so
$b_\theta=-P_1^1(\cos\theta)=\sin\theta$ and
$b_\phi=P_2^1(\cos\theta)=-3\sin\theta\cos\theta$.

\begin{table}[t!] \caption{\vspace{1mm}
Results for $\tilde{E}_\ell\tilde{B}_{\ell+1}$.
The maxima for each $\ell'$ are in bold.
}\centerline{\begin{tabular}{c|rrrrrrc}
\hline
\hline
\backslashbox{$\ell'$}{$\ell$} & $2$~~ & $4$~~ & $6$~~ & $8$~~& $10$~ & $12$~ & $\hrms$ \\
\hline
1 &{\bf4.3}&  0.0 &   0.0 &   0.0 &   0.0 &   0.0 & 1.50 \\
2 &{\bf6.5}&$-3.1$&   0.0 &   0.0 &   0.0 &   0.0 & 1.57 \\
3 &    8.3 &  3.9 &$\!\!{\bf-13.6}$&  0.0 &   0.0 &   0.0 & 1.69 \\
4 &   10.1 &  5.3 &   2.1 &$\!\!{\bf-28.0}$&  0.0 &   0.0 & 1.82 \\
5 &   12.0 &  6.5 &   4.0 &   0.0 &$\!\!{\bf-46.4}$&  0.0 & 1.95 \\
6 &   13.9 &  7.5 &   5.0 &   2.9 &${\bf-2.5}$&$\!\!{\bf-65.7}$& 1.95 \\
\hline
\label{Tab2}\end{tabular}}\end{table}

\begin{table}[t!] \caption{\vspace{1mm}
As \Tab{Tab2}, but for $\tilde{E}_\ell\tilde{B}_{\ell-1}$.
}\centerline{\begin{tabular}{c|rrrrrrc}
\hline
\hline
\backslashbox{$\ell'$}{$\ell$} & $4$~~ & $6$~~ & $8$~~& $10$~ & $12$~ & $14$~ & $\hrms$ \\
\hline
1 &~~{\bf22.6}&  0.0 &     0.0 &      0.0 &      0.0 &      0.0 &1.50 \\
2 &$  -2.1$&{\bf46.3}&     0.0 &      0.0 &      0.0 &      0.0 &1.50 \\
3 &~~  3.8 &   $-8.4$&{\bf77.9}&      0.0 &      0.0 &      0.0 &1.69 \\
4 &~~  5.7 &     1.9 &  $-16.6$&{\bf117.4}&      0.0 &      0.0 &1.82 \\
5 &~~  7.2 &     4.1 &     0.0 &$   -26.6$&{\bf165.0}&      0.0 &1.95 \\
6 &~~  8.6 &     5.4 &     2.8 &$    -2.1$&$   -38.6$&{\bf220.5}&2.08 \\
\hline
\label{Tab3}\end{tabular}}\end{table}

\begin{table}[t!] \caption{\vspace{1mm}
Similar to \Tabs{Tab2}{Tab3}, but now just for $\tilde{B}_{\ell}$.
}\centerline{\begin{tabular}{c|rrrrrrc}
\hline
\hline
\backslashbox{$\ell'$}{$\ell$} & $3$~~ & $5$~~ & $7$~~& $9$~ & $11$~ & $13$~ \\
\hline
1 &~~{\bf 5.9}&  0.0 &     0.0 &      0.0 &      0.0 &      0.0 \\
2 &$   5.9$&{\bf 8.9}&     0.0 &      0.0 &      0.0 &      0.0 \\
3 &~~  7.1 &   $ 7.3$&{\bf11.7}&      0.0 &      0.0 &      0.0 \\
4 &~~  8.6 &     7.9 &  $  8.6$&{\bf 14.6}&      0.0 &      0.0 \\
5 &~~ 10.1 &     9.0 &     8.8 &$    10.0$&{\bf 17.4}&      0.0 \\
6 &~~ 11.6 &    10.1 &     9.5 &$     9.8$&$    11.4$&{\bf 20.3}\\
\hline
\label{Tab4}\end{tabular}}\end{table}

In \Tabs{Tab2}{Tab3}, we list the two-scale polarization spectra
\EQ
K_\ell^+=\tilde{E}_\ell\tilde{B}_{\ell+1}^\ast\quad\mbox{and}\quad
K_\ell^-=\tilde{E}_\ell\tilde{B}_{\ell-1}^\ast,
\label{Kdef1}
\EN
respectively, for different values of $\ell'$.
We note here again that, because $m=0$, $\tilde{E}_\ell$ and
$\tilde{B}_{\ell\pm1}$ are real, so we can drop the asterisk.
In all cases, the integral of $h(\theta)$ over both hemispheres vanishes.
To get a sense of the strength of helicity, we therefore list in
\Tabs{Tab2}{Tab3} the rms value, $\hrms$.
We see that $\hrms$ increases only mildly with increasing values of $\ell'$.
By contrast, the extrema of $\tilde{E}_\ell\tilde{B}_{\ell+1}$ and
$\tilde{E}_\ell\tilde{B}_{\ell-1}$ increase much faster with $\ell$.
This suggests that the $\ell$-dependence of $K_\ell^-$ does not reflect
the actual $\ell$-dependence of magnetic helicity.

\Tabs{Tab2}{Tab3} also show that the maxima of both $K_\ell^+$ and
$K_\ell^-$ occur for $\ell=2(\ell'+1)$.
An exception is $\tilde{E}_\ell\tilde{B}_{\ell+1}$ for $\ell'=2$,
where the maximum still occurs at $\ell=2$.
It is important to note that the maximum of
$\tilde{E}_\ell\tilde{B}_{\ell-1}$ is much sharper in comparison to the
lower $\ell$ values than that of $\tilde{E}_\ell\tilde{B}_{\ell+1}$.
For this reason, we focus our analysis on the former quantity to
characterize the spectrum of magnetic helicity, because it serves as
the sharpest proxy of the magnetic helicity.
Also, the largest contribution to $\tilde{E}_\ell\tilde{B}_{\ell+1}$ has
the opposite sign for $\ell\geq6$.

\begin{figure*}[t!]\begin{center}
\includegraphics[width=\textwidth]{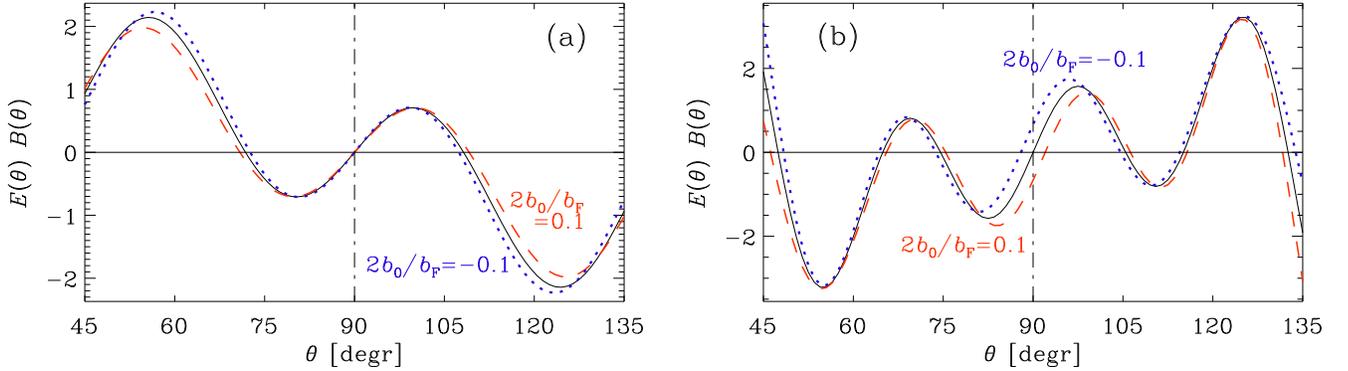}
\end{center}\caption[]{
$E(\theta)\,B(\theta)$ for $b_{\rm F}=10$ (red) and $-10$ (blue)
for (a) $l'=1$ and (b) $l'=2$.
Note that for $l'=2$, $EB\neq0$ at the equator ($\theta=90\degr$).
}\label{pfaraday}\end{figure*}

It is in principle also possible to use $\tilde{B}_{\ell,m}$ itself as
a proxy of magnetic helicity.
Its values are listed in \Tab{Tab4} for the same models as above.
We emphasize that $\tilde{B}_\ell$ has contributions only from odd values
of $\ell$.
This is because $B(\theta)$ has a dominant hemispheric $\ell=1$ variation.
By contrast, $\tilde{E}_\ell$ always has contributions for even values
of $\ell$.
This property of $\tilde{E}_\ell$ is also recovered if the field is nonhelical,
which is the case if the magnetic field is purely poloidal or purely toroidal.
On the other hand, if both are present at the same values of $\ell$,
one has helicity without hemispheric modulation.
In that case, $\tilde{B}_\ell$ has contributions only from even values
of $\ell$, while $\tilde{E}_\ell$ vanishes.

\subsection{Analogy with Faraday-rotated fields}

\cite{SF97} calculated the $B$ mode polarization of the cosmic microwave
background radiation in the presence of a uniform magnetic field and found
correlations between the temperature at spherical harmonic degree $\ell$
and the $B$ mode at degrees $\ell+1$ and $\ell-1$; see also \cite{SHM04}.
Such constructs are reminiscent of those in \Eq{Kdef1}.
In their case, the uniform magnetic field led to a superposition of
Faraday-rotated fields with different angles over the depth near the
last scattering surface.

\begin{figure*}[t!]\begin{center}
\includegraphics[width=\textwidth]{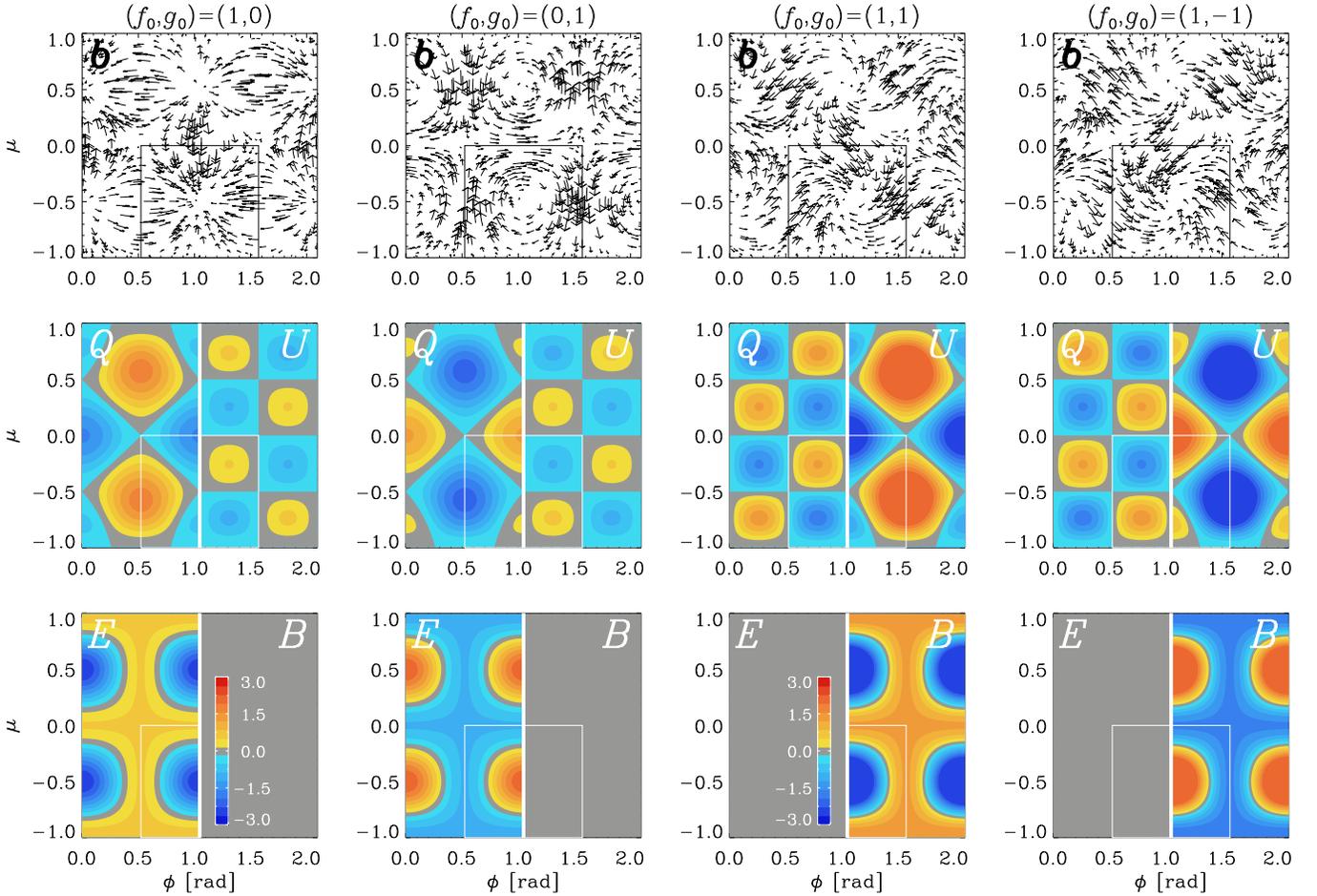}
\end{center}\caption[]{
$\bb(\phi,\mu)$ vectors compared with split representations
of $(Q,U)$ and $(E,B)$ for the four combinations $(f_0,g_0)=(1,0)$,
$(0,1)$, and $(1,\pm1)$ with $\ell=4$ and $m=3$.
Individual cross, ring, and swirl-like patterns are highlighted by squares,
along with their positions in the $(Q,U)$ and $(E,B)$ diagrams.
}\label{pcombined_f1m1}\end{figure*}

An analogy with Faraday rotation is indeed justified, because both
magnetic helicity and Faraday rotation lead to similar effects that,
in combination, can either enhance or diminish the resulting polarized
intensity \citep{Soko98,BS14,HF14}.
The presence of magnetic helicity leads either to a correlation or an
anticorrelation between the rotation measure and the total polarized
intensity \citep{VS10}, depending on whether one looks along or against
the direction of the uniform magnetic field.
This explains the analogy with the present case, where we have opposite
signs of magnetic helicity in the two hemispheres.

To demonstrate the effect of Faraday rotation in the present context,
we now include the radial magnetic field.
In fact, the poloidal field associated with the latitudinal component
$b_\theta=-b_0 P_{\ell'}^1(\cos\theta)$ of our earlier examples implies
\EQ
b_r=(\ell'+1)\,b_0 P_{\ell'}(\cos\theta),
\EN
where $\dd[\sin\theta P_{\ell'}^1(\cos\theta)]/\dd\cos\theta=
-\ell'(\ell'+1)P_{\ell'}(\cos\theta)$ has been used, and the
$\ell'+1$ factor follows from \Eq{aellbell} and the fact that
$-\ell\tilde{a}_\ell/R=b_0$.
We consider models with $\ell'=1$ and $2$.
Faraday rotation rotates the phase angle of the complex polarization,
so \Eq{Emissivity} has to be replaced by
\EQ
p=-\epsilon\,(b_\theta+\ii b_\phi)^2\,e^{2\ii b_r/b_{\rm F}},
\EN
where $b_{\rm F}=(k_{\rm F}n_{\rm e}\lambda^2 d)^{-1}$,
with $k_{\rm F}=2.6\times10^{-17}\G^{-1}$ being a constant
\citep[e.g.][]{ACD94}, $n_{\rm e}$ the mean electron density,
$\lambda$ the wavelength, and $d$ the geometrical depth.
For example, for $n_{\rm e}=10^{14}\cm^{-3}$,
$\lambda=600\nm$, and $d=100\km$, we have $b_{\rm F}\approx10\kG$.
Since the actual surface magnetic field is much weaker, Faraday rotation
will only be a small effect as far as the average field is concerned.
However, given that the effect is highly nonlinear, it is usually not
negligible in active regions and sunspots.

To assess the effects of Faraday rotation on the resulting $EB$
correlation, it is instructive to look at the latitudinal dependence
of the product $E(\theta)\,B(\theta)$ for two representative cases:
one where $\ell'$ is odd and one where it is even.
The result is shown in \Fig{pfaraday} for $\ell'=1$ and $2$, using
$b_0/b_{\rm F}=\pm0.1$ and comparing with the case without Faraday rotation.
For clarity, we only show the range $45\degr\leq\theta\leq135\degr$.
For the Sun, as alluded to above, the actual values of $b_0/b_{\rm F}$
will be much smaller and the Faraday rotation effect hardly noticeable
for the average field.

We see that for $\ell'=1$, Faraday rotation causes an enhancement
(reduction) of the helicity-induced $EB$ correlation if $b_{\rm F}$
is negative (positive); see \Figp{pfaraday}{a}.
This agrees qualitatively with the result of \cite{SF97}, because a
uniform magnetic field corresponds to an odd value $\ell'$.

For $\ell'=2$, on the other hand, we have a mixed hemispheric dependence
of $EB$ with finite values at the equator.
In the case of the Sun, of course, the large-scale magnetic field has
odd symmetry around the equator.
This also applies to the field within sunspots.
The difference between leading and following sunspots would weaken the
net effect, but not its systematic north--south dependence.
We can therefore conclude that Faraday rotation does not compromise the
ability to detect magnetic helicity from $EB$, provided the Faraday
effect remains subdominant compared with the helicity effect, i.e.,
$\lambda$ is small enough.

\section{Nonaxisymmetric examples}
\label{NonAxisymmetric}

\subsection{Two-dimensional patterns of $E$ and $B$}

We now consider two-dimensional examples in the $(\phi,\mu)$ plane,
where $\mu=\cos\theta$.
Analogous to earlier work, we consider the magnetic field
$(b_\phi,b_\mu)\equiv(b_\phi,-b_\theta)$ to be given by $\bb=\FF+\GG$,
where
\EQ
F_i=\nabla_i f\quad\mbox{and}\quad
G_i=\epsilon_{ij}\nabla_j g,
\label{FGformulation}
\EN
using
\EQ
f=-f_0 Y_{\ell m} \quad\mbox{and}\quad
g=g_0 Y_{\ell m}.
\EN
The complex linear polarization is then computed as
$p=-(b_\theta+\ii b_\phi)^2=(b_\phi-\ii b_\theta)^2
=(b_\phi+\ii b_\mu)^2$.
Following BBKMRPS, we consider four combinations, namely
$(f_0,g_0)=(1,0)$, $(0,1)$, and $(1,\pm1)$.
In \Fig{pcombined_f1m1}, we show the result for $\ell=4$ and $m=3$.
All quantities are plotted as a function of $\phi$ and $\mu=\cos\theta$.
This corresponds to the Lambert azimuthal equal-area projection.
We recover familiar structures corresponding to a star-like and
ring-like features for negative and positive $E$ polarizations
and swirly inward clockwise and counter-clockwise patterns
for negative and positive $B$ polarizations.
These structures agree with those in Figure~2 of BBKMRPS.
We recall, however, that we follow here the sign convention
of \cite{Dur08}, in which our \Eq{EBfromQU} becomes
$\tilde{R}(k_x,k_y)=-(\hat{k}_x-\ii\hat{k}_y)^2\tilde{p}(k_x,k_y)$
in the Cartesian limit, 
where $k_x$ and $k_y$ are the components of the two-dimensional
wavevector and hats indicate unit vectors.
Equation~(3) of BBKMRPS followed the sign convention of \cite{ZS97},
but their Figure~2 showed polarization vectors, which are at right
angles to the magnetic field vectors, giving therefore the same
orientation as the magnetic vectors in the Durrer convention shown in
our \Fig{pcombined_f1m1}.

\begin{figure*}[t!]\begin{center}
\includegraphics[width=\textwidth]{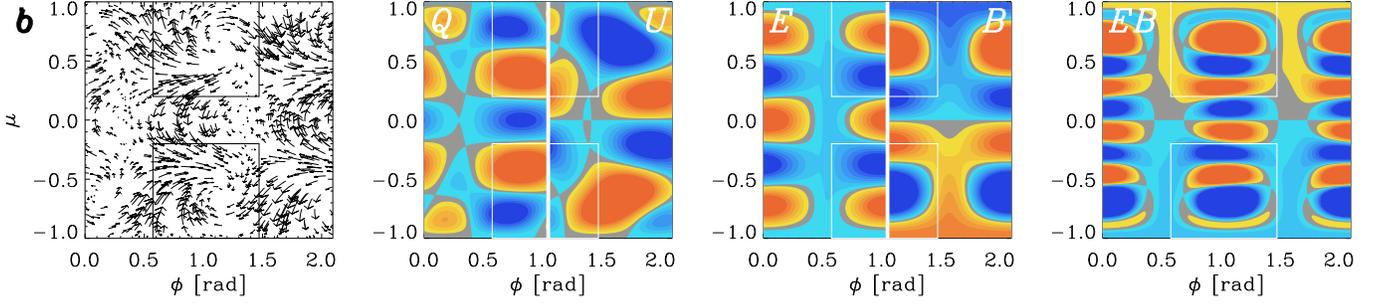}
\end{center}\caption[]{
$\bb(\phi,\mu)$ vectors compared with split representations of $(Q,U)$
and $(E,B)$, and a representation of the product $EB$, for the model of
\Sec{HemisphericHelicityModulation} with $\ell=4$ and $m=3$.
Note the opposite sense of swirl of eddies in the northern and southern
hemispheres, as highlighted by the squares.
}\label{pcombined_Dl1}\end{figure*}

\subsection{Formulation in terms of superpotentials}

Nonaxisymmetric magnetic fields can no longer be expressed in a form
analogous to \Eq{axisymb}, but we must instead employ the superpotentials
$S$ and $T$ in the form
\EQ
\bb=\nab\times\nab\times(\rr S)+\nab\times(\rr T).
\EN
The first part corresponds to the poloidal field and the second to the
toroidal field.
The two superpotentials are expanded in terms of spherical harmonics, so
\EQ
(S,T)(\theta,\phi)=\sum_{\ell=1}^{N_\ell}\sum_{m=-\ell}^{\ell}
(S_{\ell m},T_{\ell m})\,Y_{\ell m}(\theta,\phi),
\EN
with the inverse transformation given by
\EQ
(\tilde{S}_{\ell m},\tilde{T}_{\ell m})=\int_{4\pi}
(S,T)(\theta,\phi)\,Y_{\ell m}^\ast(\theta,\phi)\,
\sin\theta\,\dd\theta\,\dd\phi.
\EN
As in \Sec{RepresentationAxisymmetric}, we assume that the radial
dependence of $\tilde{S}_{\ell m}(r)$ is proportional to $r^{-(\ell+1)}$.
This implies that
\EQ
\frac{\partial}{\partial r}(r\tilde{S}_{\ell m})=
-\ell \tilde{S}_{\ell m} \quad\mbox{(for $r=R$)}.
\EN
For chosen values of $\ell$ and $m$, we can then write
\EQ
b_\theta(\theta,\phi)=\Rey\left(-\ell\tilde{S}_{\ell m}\nabla_\theta Y_{\ell m}
+\tilde{T}_{\ell m}\nabla_\phi Y_{\ell m}\right),
\label{bthetaRep}
\EN
\EQ
b_\phi(\theta,\phi)=\Rey\left(-\ell \tilde{S}_{\ell m}\nabla_\phi Y_{\ell m}
-\tilde{T}_{\ell m}\nabla_\theta Y_{\ell m}\right).
\label{bphiRep}
\EN
Note in this connection that for axisymmetric models, $b_\theta$ and
$b_\phi$ are related to $Y_{\ell m}(\theta,\phi)$ via $\theta$ derivatives.
This shows that the reason for having expanded $a_\phi(\theta)$ and
$b_\phi(\theta)$ in \Eq{axisymbExpand} in terms of $P_\ell^1(\cos\theta)$
is that the $\theta$ derivative of the Legendre polynomials gives
$\dd P_\ell(\cos\theta)/\dd\theta=P_\ell^1(\cos\theta)$.
Analogously to the axisymmetric case, we choose
$\tilde{T}_{\ell m}=-\ell\tilde{S}_{\ell m}/R=b_0R$.

The formulation given by \Eqs{bthetaRep}{bphiRep} agrees with that given
by \Eq{FGformulation}, provided we replace
\EQ
f\to -\ell \tilde{S}_{\ell m} Y_{\ell m}(\theta,\phi),\quad
g\to \tilde{T}_{\ell m} Y_{\ell m}(\theta,\phi).
\EN
This formulation suggests that the nonaxisymmetric generalization
of \Eq{HelField} is given by
\EQ
f\to -\ell' \tilde{S}_{\ell' m'} Y_{\ell' m'} R,\quad
g\to \tilde{T}_{\ell'+1\, m'} Y_{\ell'+1\, m'},
\EN
and that
\EQ
H_{\ell'}^\pm=\sum_{m'=-\ell'}^{\ell'} 2\ell'(\ell'+1)
\tilde{S}_{\ell' m'} \tilde{T}_{\ell'\pm1\, m'}^\ast
\label{Hpmdef}
\EN
can be used as a global two-scale measure of the magnetic helicity spectrum.
In the following, we use $H_{\ell'm'}^+$ to specify the amplitude of a
single mode; $H_{\ell'm'}^-$, by contrast, vanishes in our single-mode
examples by construction.
We also use $H_\ell^+$ for solar magnetograms.

\subsection{Hemispheric helicity modulation}
\label{HemisphericHelicityModulation}

In the examples considered above, either $E$ or $B$ was zero; see the
gray sub-panels in the split representation of \Fig{pcombined_f1m1}.
We now consider examples where both are nonvanishing.
Specifically, we reconstruct examples where
\EQ
K_{\ell}^\pm\equiv\sum_{m=-\ell}^\ell
\tilde{E}_{\ell m}\tilde{B}_{\ell\pm1\, m}^\ast
\label{KellmpmDef}
\EN
is nonvanishing.
As noted in the previous section, we do this by using fields where
\EQ
-\ell' \tilde{S}_{\ell' m'}/R=\tilde{T}_{\ell'+1\, m'}=-b_0
\EN
is a constant for fixed $\ell'$ and $m'$.
This is equivalent to our choice
$-\ell'\tilde{a}_{\ell'}/R=\tilde{b}_{\ell'+1}=b_0$ in \Sec{Spin2}.
Furthermore, the models of \Fig{p2_4panels} correspond to
$(f_0,g_0)=(1,-1)\times4\pi/\sqrt{(2\ell+1)(2\ell+3)}$.
The result is shown in \Fig{pcombined_Dl1}, again for $\ell'=4$ and $m'=3$.
We see that $E$ is always symmetric about the equator and $B$ is
antisymmetric about the equator.
The product $EB$ is therefore antisymmetric about the equator,
which reflects the opposite signs of magnetic helicity in the
two hemispheres.

The last panel of \Fig{pcombined_Dl1} shows that, although the product
$EB$ is mostly positive in the north and negative in the south, there
are also extended regions of opposite sign.
Quantitatively, we find that $2\bra{EB}/\bra{E^2+B^2}=\pm0.25$, where
the upper (lower) sign applies to the northern (southern) hemisphere.

\begin{table}[t!] \caption{\vspace{1mm}
Values of $\tilde{E}_{\ell m}$, $\tilde{B}_{\ell-1\,m}$, and
$\tilde{E}_{\ell m}\tilde{B}_{\ell-1\,m}$ for $\ell'=4$ and $m'=3$.
}\centerline{\begin{tabular}{c|ccccc}
$\ell$ &   $2$  &    $4$  &    $6$          & $8$             &   $10$  \\
\hline
$\tilde{E}_{\ell\;0}$    &  $1.00$ & $-0.37$ & $-2.61$         &  $2.43$         &$\!\!\!\!\!-0.63$ \\
$\tilde{B}_{\ell-1\;0}$ &   $0$   & $-2.37$ & $-3.14$         & $ 3.25$         &$\!\!\!\!\!-0.82$ \\
$\tilde{E}_{\ell 0}\tilde{B}_{\ell-1\,0}^\ast$
                        &   $0$   &$\,\;\;0.89$&$\,\;\;{\bf8.19}$& $ {\bf7.88}$ & $ 0.51$ \\
\hline
$\tilde{E}_{\ell\,6}$    & $  0 $  &  $  0 $ &$\!\!0.33-0.52\ii$&$-0.19-0.15\ii$ &  $1.94$ \\
$\tilde{B}_{\ell-1\;6}$ &   $0$   &  $  0 $ &  $  0 $         &  $1.67$         & $ 3.18$ \\
$\tilde{E}_{\ell 6}\tilde{B}_{\ell-1\,6}^\ast$
                        &   $0$   &  $  0 $ &  $  0 $         & $-0.31-0.24\ii$ & $ {\bf6.19}$ \\
\hline
$K_\ell^-$                     &   $0$& $0.89$ & $ 8.19$ &  $7.26$ &$\!\!\!{\bf12.89}$ \\
$\sum|\tilde{E}_{\ell m}|^2$   &$1.00$& $0.14$ & $ 7.57$ &  $6.02$ &$7.92$ \\
$\!\!\sum|\tilde{B}_{\ell-1\,m}|^2\!\!$&$   0$& $5.62$ & $ 9.86$ &$\!\!\!\!\!\!16.1$&$\!\!\!\!20.90$ \\
$c_{\rm A}(\ell)$              &   $0$& $0.31$ & $ 0.94$ &  $0.66$ & $ 0.89$
\label{TabX}\end{tabular}}\end{table}

\begin{figure*}[t!]\begin{center}
\includegraphics[width=\textwidth]{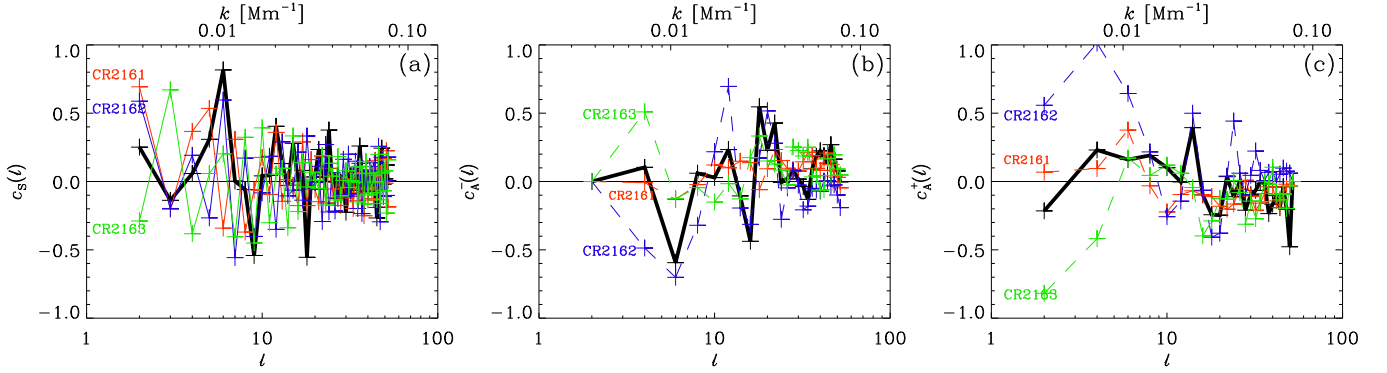}
\end{center}\caption[]{
(a) $c_{\rm S}(\ell)$, (b) $c_{\rm A}^-(\ell)$, and
(c) $c_{\rm A}^+(\ell)$ for the full data set covering CRs~2161--2163
(broad solid lines), compared with the corresponding individual results
for CRs~2161 (red), 2162 (blue), and 2163 (green).
}\label{pKm}\end{figure*}

In \Tab{TabX}, we list all nonvanishing coefficients
$\tilde{E}_{\ell m}$ and $\tilde{B}_{\ell m}$ for
our example with $\ell'=4$ and $m'=3$.
For $m\neq0$, the only nonvanishing contributions come from $m=\pm6$.
Note also that $\tilde{E}_{\ell m}$ is now complex,
while all other coefficients are still real.
The dominant contributions to the parity-odd correlation
come from the product $\tilde{E}_{\ell m}\tilde{B}_{\ell-1\,m}$
with $\ell=2(\ell'-1)=6$ and $\ell=2\ell'=8$ for $m=0$, and
$\ell=2(\ell'+1)=10$ for $m=2m'=6$.

\section{Solar synoptic vector magnetograms}
\label{Applications}

\subsection{Spectra of global two-scale helicity proxies}
\label{GlobalProxies}

We now apply the global two-scale approach to the same solar synoptic
vector magnetograms that were studied by BPS using the semi-global
approach.
As alluded to in the introduction, we use ``$\pi$-ambiguated''
magnetic fields expressed in terms of pseudo-polarization data.
Thus, we only utilize the two horizontal components, $b_\theta$
and $b_\phi$, to compute the complex linear polarization
$p(\theta,\phi)=-(b_\theta+\ii b_\phi)^2$.
The emissivity prefactor in \Eq{Emissivity} has been set to unity
because, in the following, we only work with normalized spectra.
We then compute $\tilde{E}_{\ell m}$ and $\tilde{B}_{\ell m}$
and study the spectra $K_\ell^\pm$; see \Eq{KellmpmDef}.
We normalize them analogously to those in BBKMRPS and write them as
\EQ
c_{\rm A}^\pm(\ell)=\frac
{\sum_{m=-\ell}^{\ell} 2\tilde{E}_{\ell m}\tilde{B}_{\ell\pm1\, m}^\ast}
{\sum_{m=-\ell}^{\ell}\left(|\tilde{E}_{\ell m}|^2+|\tilde{B}_{\ell-1\, m}|^2\right)}.
\EN
Because we sum over positive and negative $m$, the values of
$c_{\rm A}^\pm(\ell)$ are aways real.
They vary between $-1$ and $+1$.
We recall that, based on the comparison of \Tabs{Tab2}{Tab3} in
\Sec{Spin2}, we expect $c_{\rm A}^-(\ell)$ to be a better proxy of
magnetic helicity than $c_{\rm A}^+(\ell)$.

\begin{figure*}[t!]\begin{center}
\includegraphics[width=\textwidth]{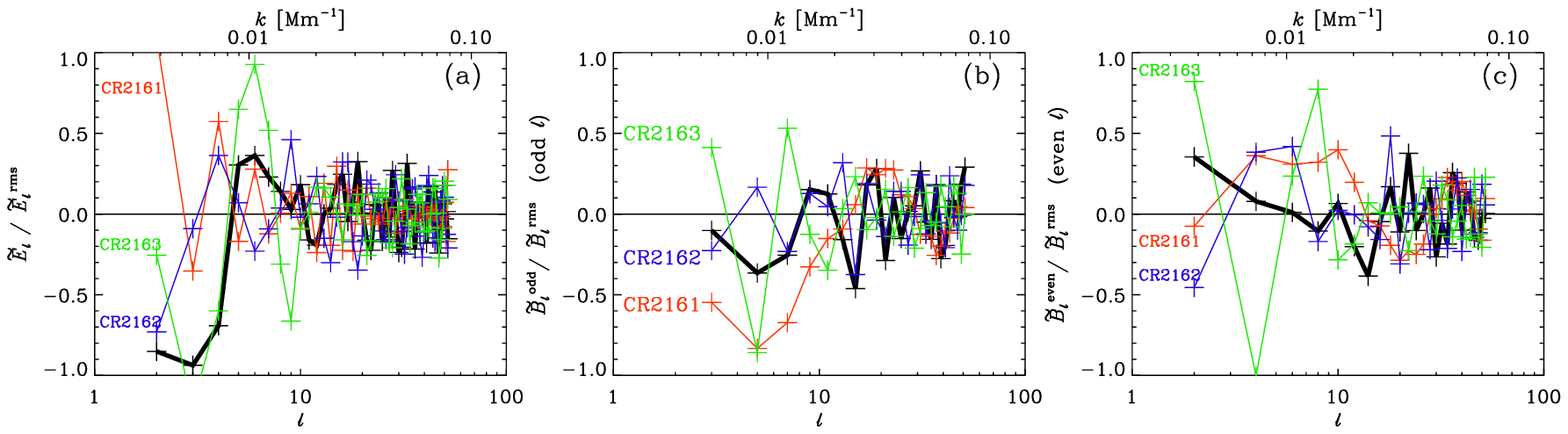}
\end{center}\caption[]{
(a) $\tilde{E}_\ell/\tilde{E}_\ell^{\rm rms}$ (for even and odd $\ell$),
(b) $\tilde{B}_\ell^{\rm odd}/\tilde{B}_\ell^{\rm rms}$ (only for odd values
of $\ell$), and (c) $\tilde{B}_\ell^{\rm even}/\tilde{B}_\ell^{\rm rms}$
(for even values of $\ell$), for the full data set covering CRs~2161--2163
(broad solid lines), compared with the corresponding individual results
for CRs~2161 (red), 2162 (blue), and 2163 (green).
}\label{pBm}\end{figure*}

Following BBKMRPS, we also compute the normalized difference of the spectra
of $EE$ and $BB$ polarizations as
\EQ
c_{\rm S}(\ell)=\frac
{\sum_{m=-\ell}^\ell\left(|\tilde{E}_{\ell m}|^2-|\tilde{B}_{\ell m}|^2\right)}
{\sum_{m=-\ell}^\ell\left(|\tilde{E}_{\ell m}|^2+|\tilde{B}_{\ell m}|^2\right)}.
\EN
This quantity varies between $-1$ and $+1$.
It vanishes when the $EE$ and $BB$ polarizations have the same amplitude,
and it is $1/3$ if the amplitude of the $EE$ polarization is twice
that of the $BB$ polarization, as was found in recent dust foreground
measurements of the interstellar medium \citep{Adam16,Akrami18}.
To facilitate comparison with earlier work, we define
\EQ
L^2=\ell(\ell+1),
\label{L2def}
\EN
and plot $c_{\rm S}$ and $c_{\rm A}^\pm$ also versus
the approximate wavenumber $k=L/R$.
As in BPS, we use the combined synoptic vector magnetograms
of three CRs, 2161, 2162, and 2163.
They are based on the full-disk vector magnetograms obtained
from the Helioseismic and Magnetic Imager on board the {\em
Solar Dynamics Observatory} and have been processed by Yang
Liu\footnote{\url{http://hmi.stanford.edu/hminuggets/?p=1689}}
(Stanford).

In \Fig{pKm}, we show $c_{\rm S}(\ell)$ and $c_{\rm A}^\pm(\ell)$ both
for the full data set of all three CRs and also separately for CRs~2161,
2162, and 2163.
For the full data set, the total azimuthal angle is $6\pi$, and the
integration in \Eq{EBfromQU} is carried out over $12\pi$ instead of $4\pi$.
Similar to our earlier semi-global analysis, $c_{\rm S}$ shows large
variations, but is mostly positive for $\ell\leq10$, corresponding to
the wavenumber $k=L/R\leq0.014\Mm^{-1}$.
Furthermore, $c_{\rm A}^-$ shows negative values for similar $\ell$,
while $c_{\rm A}^+$ has the opposite sign, which is in agreement
with our expectations based on the comparison of \Tabs{Tab2}{Tab3}.
For larger $\ell$, both $c_{\rm A}^+$ and $c_{\rm A}^-$ are again very
noisy, although $c_{\rm A}^-$ may be mostly positive, while $c_{\rm A}^+$
may be mostly negative.

To have an estimate of the uncertainty of our results, we also plot the
spectra separately for each of the three CRs.
These results are broadly consistent with those of the full data set.
The tendency of obtaining positive values of $c_{\rm S}$ at $\ell<10$
is also seen individually for all three CRs.
By contrast, the tendency of obtaining negative values of $c_{\rm A}^-$ for
$\ell<10$ is seen for CRs~2161 and 2162, but not for CR~2163 at $\ell=4$.
However, for $\ell=6$, all three data sets give the same (negative) sign
of $c_{\rm A}^-$.

As noted before, $\tilde{B}_\ell$ can itself be used as a helicity proxy,
so we now determine it for the same three CRs.
For completeness, we also analyze $\tilde{E}_\ell$ in a similar fashion.
Owing to nonaxisymmetry, we have contributions from different values
of $m$.
It is then useful to define
\EQ
\tilde{B}_\ell=\sum_{m=-\ell}^{\ell}\tilde{B}_{\ell\,m},\quad
\tilde{B}_\ell^{(2)}=\sum_{m=-\ell}^{\ell}|\tilde{B}_{\ell\,m}|^2.
\EN
In the following, we plot $\tilde{B}_\ell$ and the ratio
$\tilde{B}_\ell/\tilde{B}_\ell^{\rm rms}$, where
$\tilde{B}_\ell^{\rm rms}=[\tilde{B}_\ell^{(2)}]^{1/2}$ is the
rms value.
We define $\tilde{E}_\ell$ and the ratio
$\tilde{E}_\ell/\tilde{E}_\ell^{\rm rms}$ analogously.
For $\tilde{B}_\ell$, we only expect to see a hemispheric modulation
for odd values of $\ell$.
Therefore, to distinguish the contributions from odd and even
values of $\ell$, we denote them as $\tilde{B}_\ell^{\rm odd}$
and $\tilde{B}_\ell^{\rm even}$.
The results are shown in \Fig{pBm} as a function of $\ell$.
We see that both $\tilde{E}_\ell$ and $\tilde{B}_\ell^{\rm odd}$ are
negative for small values of $\ell$, while $\tilde{B}_\ell^{\rm even}$
is positive.

The fact that $\tilde{E}_\ell$ is mostly negative for $\ell<5$ suggests
that, on large length scales, the magnetic field structures are mostly
star-like, but in the range $5<\ell<10$, they are mostly ring-like.
However, no direct visual evidence of this has been reported as yet.
For the $B$ polarization, on the other hand, the negative values
for odd $\ell$, i.e., for $\tilde{B}_\ell^{\rm odd}$, may
reflect a positive magnetic helicity on large length scales; see
\Sec{HemisphericHelicityModulation}.
This agrees with the negative sign found for $c_{\rm A}^-$.
Moreover, as seen in \Tab{Tab2}, $K_\ell^+$ tends to have the opposite
sign.
This agrees with what is found for $c_{\rm A}^+$ in \Figp{pKm}{c}.

\subsection{Spectra of the global two-scale magnetic helicity}
\label{GlobalHelicity}

Finally, we consider $H_\ell^\pm$.
We normalize it by the solar radius $R$, which is set to unity in our
work, so we plot here the ratio $H_\ell^\pm/R$, which has units of $\G^2$;
see \Eq{Hpmdef} for the definition.
To obtain $\tilde{S}_{\ell m}$, we use the observed radial magnetic field
component, $b_r$, compute the spherical harmonics decomposition to find
$\tilde{b}_{r,\ell m}$ and thus $\tilde{S}_{\ell m}=\tilde{b}_{r,\ell m}/L^2$;
see \Eq{L2def}.
Analogously, we compute $\tilde{T}_{\ell m}$ from the radial component
of the current density, $j_r$.
For the vector magnetograms, the components of the magnetic field are
given in uniform intervals of $\mu=\cos\theta$.
We therefore write the radial component of $\jj=\nab\times\bb$ as
\EQ
j_r=\cot\theta\,b_\phi-\sin\theta\,{\partial b_\phi\over\partial\mu}
-{1\over\sin\theta}\,{\partial b_\theta\over\partial\phi}.
\EN
We then compute the spherical harmonics decomposition
to find $\tilde{j}_{r,\ell m}$ and thus compute
$\tilde{T}_{\ell m}=\tilde{j}_{r,\ell m}/L^2$.

\begin{figure}[t!]\begin{center}
\includegraphics[width=\columnwidth]{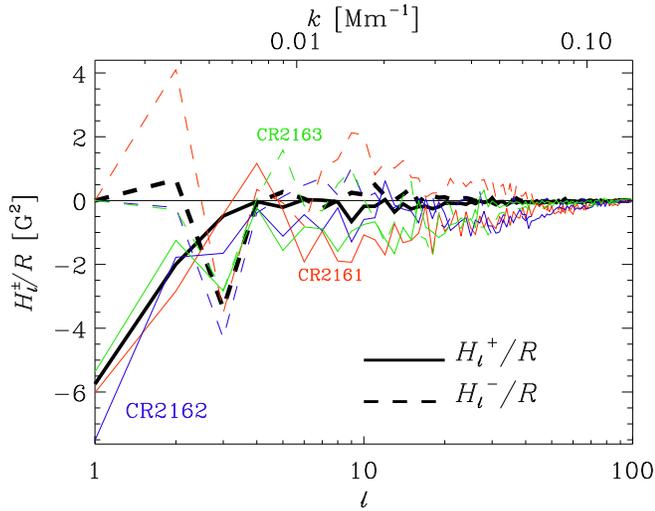}
\end{center}\caption[]{
$H_\ell^\pm/R$ versus $\ell$ for the full data set covering CRs~2161--2163
(broad solid lines), compared with the corresponding individual results
for CRs~2161 (red), 2162 (blue), and 2163 (green).
The solid (dashed) lines give the results for odd (even) values of $\ell$.
}\label{pHp}\end{figure}

In \Fig{pHp} we plot $H_\ell^\pm/R$ versus $\ell$.
We see that $H_\ell^+$ and $H_\ell^-$ are negative for most values
of $\ell$.
Thus, there is no clear evidence for a positive magnetic helicity at large
length scales.
This is surprising in view of the previous findings based on the $B$
polarization that did suggest positive magnetic helicity on large
length scales.
Of course, previous work has long shown negative magnetic helicity in the
northern hemisphere and positive in the southern \citep{See90,PCM95},
including the work of BPS.
It may therefore indeed be true that there is no sign change in
$H_\ell^\pm$ at the photosphere, and that the sign change in the helicity
proxies may reflect physical properties of the field at some layer above
the photosphere.
However, it could also be an effect of the phase within the solar cycle,
as suggested by \cite{Singh18}, which could then also explain the evidence
for a bihelical field found by \cite{PP14}.
Systematic cycle related magnetic helicity variations are indeed well
documented \citep{KKMRSZ03,ZSPGXSK10,PPLK19}.

It also is surprising that $H_\ell^\pm$ is, for all CRs, consistently
much larger at $\ell=1$ than for any other value of $\ell$.
In BPS, by contrast, we found a rapid decline of power for $0.01\Mm^{-1}$;
see Figure~8 therein, but that work was based on a semi-global approach
which is unable to recover the low $k$ values correctly.
Conversely, it is possible that the global approach overemphasizes the
polar fields.
This may be a concern mainly for the $E$ and $B$ polarization.
Indeed, looking at \Fig{p2_4panels}, we see that the clearest hemispheric
dependence in $B$ is seen at the poles, while at lower latitudes, $E$
and $B$ have no definite correlation.
This may well be a general problem with the $EB$ approach that
should be clarified studying the signs of $E$ and $B$ locally.
It would be important to assess the statistical robustness of these
results by inspecting the magnetic helicity signatures for many more CRs.

\section{Implications for dynamo theory}
\label{Implications}

The $\alpha$ effect in dynamo theory is the main candidate for explaining
the production of large-scale magnetic fields in the Sun.
One of its signatures is the production of magnetic helicity of opposite
signs.
Such a magnetic field is called bihelical.
Figures~\ref{pKm} and \ref{pBm} present some support for this
assertion, in addition to the earlier results of \cite{PP14} and
\cite{Singh18}.
Our inspection of $H_\ell^+$ and $H_\ell^-$ does not support this,
however.
Whether this is indeed related to potential problems regarding the
$\pi$ ambiguity is, however, unclear, and one would like to see more
evidence before continuing to speculate further on this.
There is, however, the possibly that it might be related to the
anticipated sign reversal of magnetic helicity some small distance above
the photosphere.
We elaborate on this possibility next. 

To put the various findings into a broader perspective, it is important to
realize that in the solar wind, far away from the solar dynamo, evidence
for a bihelical magnetic field has also been presented \citep{BSBG11}.
However, the sign of magnetic helicity is at all wavenumbers opposite
to what it is at the solar surface.
This was then found to be a generic phenomenon of any system consisting
of a dynamo region adjacent to a nondynamo region; see the work of
\cite{WBM11,WBM12} of a turbulent dynamo simulation with a simple
stellar corona, and the earlier work of \cite{BCC09} in the context of
galactic halos.
We do not know exactly where the sign would swap.
It has been suggested that it could occur in the lower corona,
where the plasma beta crosses unity \citep{BSB18}.
This could be detectable by measuring in situ polarized emission from
within the corona \citep{BAJ17}.
On the other hand, if it happened in the chromosphere, in layers
accessible to a direct face-on measurement of the $EB$ cross-correlation,
this sign change might be detectable using the method discussed in the
present paper.

A major difficulty in detecting an overall sign change of handedness
through the $EB$ cross-correlation lies in the fact that the $E$
polarization is strongly associated with the magnetic field topology.
This particular property could be characterized, for example, by its
correlation with temperature $T$ (related to the intensity or Stokes $I$).
This is a parity-even correlation, which can have either sign, and it
may be this quantity, in addition to $EB$, that also shows a systematic
variation with height.
Not much is known about this, except that in the dust polarization of
the Galactic foreground, the $ET$ correlation is known to be positive
\citep{Akrami18}.
We also know that the $E$ polarization is highly skewed and its skewness
depends systematically on the physics governing the magnetic field.
Ambipolar diffusion, for example, is known to affect the skewness of
$E$ in a systematic way \citep[see Figure~13 of][]{Bra19AD}.
This is also reflected in the fact that the $EE$ correlation can be
different from the $BB$ correlation, i.e., $c_{\rm S}\neq0$, as has been
found in the present work; see \Figp{pKm}{a}.
Addressing these new questions raised above is of direct relevance to
assessing the possibility of a radial sign reversal of the magnetic
helicity, as predicted by dynamo theory and as has been found from
magnetic helicity measurements in the solar wind.

\section{Conclusions}
\label{Concl}

This work has addressed two critical issues in the calculation of a
proxy of solar magnetic helicity spectra: the $\pi$ ambiguity and the
systematic north--south sign change of magnetic helicity.
The problem of the $\pi$ ambiguity has been addressed previously
(BBKMRPS) by calculating the $EB$ cross--correlation from local
Cartesian patches.
This quantity was shown to be a proxy of magnetic helicity under
inhomogeneous conditions, in particular for rotating stratified
convection.
The problem of the systematic north--south variation has also been
addressed previously, but only in a semi-global fashion; see BPS.
Here, we have generalized this approach to a fully global one by first
calculating the parity-even and parity-odd $E$ and $B$ polarizations
globally using spin-2 spherical harmonics, and then correlating them at
spherical harmonic degrees that are shifted by one relative to the other.
This approach is analogous to what was done in the semi-global Cartesian
approach of BPS.
However, unlike their formalism, the present one is heuristic and
has not been derived rigorously from a correlation function that depends
on mean and relative coordinates; see \cite{RS75}.
It is not entirely obvious that this is even possible but, if it is,
the result may well look similar to what has been proposed here.
Through the examples constructed here, we have demonstrated that the correlation
$\tilde{E}_{\ell m}\tilde{B}_{\ell-1\, m}^\ast$ can act as a proxy
of the magnetic helicity, which itself is characterized globally by the product
$\tilde{S}_{\ell m}\tilde{T}_{\ell+1\, m}^\ast$.

In the quest for finding clear evidence of an opposite sign of magnetic
helicity at large length scales, one has to tackle the problem of the
$\pi$ ambiguity in the weak-field regions that occupy the majority of
the solar surface.
A standard approach to $\pi$ disambiguation in those regions is the
random disambiguation, which is problematic and may have been responsible
for the relatively low spectral power at wavenumbers around and below
$0.03\Mm^{-1}$ \citep{Singh18} and also for what looked like a random
sign in the resulting magnetic helicity at those wavenumbers.
In fact, the present results now suggest that there is maximum power
at the very smallest wavenumbers around and below $0.01\Mm^{-1}$.

Our results show that, in the northern hemisphere, where the small-scale
magnetic helicity is negative, $\tilde{E}_{\ell m}\tilde{B}_{\ell-1\, m}^\ast$
is positive.
Likewise, the large-scale field is expected to have positive magnetic
helicity in the northern hemisphere and
$\tilde{E}_{\ell m}\tilde{B}_{\ell-1\,m}^\ast$ is now found to be negative.
Thus, our proxy has the opposite sign to the magnetic helicity.
This agrees with what was found based on the numerical
simulations of BBKMRPS.
This result is not, however, based on the actual helicity $H_\ell^\pm$,
but rather on the helicity proxy.
As mentioned in \Sec{GlobalHelicity}, there could be a general difficulty
with the $EB$ approach in that its highest sensitivity is at the poles.
At lower latitudes, the method suffers a significant amount of
cancellation, as can be anticipated from \Fig{p2_4panels} for $\ell=4$.

Regarding the absence of a clear $EB$ signal in the analysis of solar
$Q$ and $U$ polarization in the work of BBKMRPS, it should be noted
that their results are much more noisy, although in hindsight not so
dissimilar from the present ones.
Tentatively, they found values at small and large length scales that
agree with those here: positive $c_{\rm S}(k)$ at $k=0.01\Mm^{-1}$
along with $c_{\rm A}(k)$ at similar values of $k$.
However, the main reason for their noisy result lies probably in the
fact that their linear polarization data were too contaminated by other
factors, as was already discussed in BBKMRPS.

The present approach of computing the $EB$ signal from the magnetic
field rather than the observed polarization combines the best aspects
of two worlds.
It uses the elaborate inversion technique of spectropolarimetry to obtain
the magnetic field, but is insensitive to the problems associated with
the $\pi$ ambiguity.
What is perhaps unsatisfactory, however, is the fact that the
line-of-sight magnetic field ($b_\|$) or the circular polarization are
not used in the present approach.
No corresponding idea has yet been proposed that would combine these
two pieces of information.
Simply correlating $b_\|$ with $E$ or $B$ may not yield anything useful
because in simple patterns such as those of \Fig{pcombined_f1m1}, the
wavelength of $b_\|$ is always twice that of $E$ or $B$, so it would
lead to a cancellation.
This is because $E$ and $B$ are related to the square of the magnetic
field.
Therefore, the spatial wavelengths of the $E$ and $B$ patterns would
agree with that of $b_\|^2$, but then the potentially useful information
implied by the sign of $b_\|$ is lost.
So, it is not obvious what to do with $b_\|$ in this context.

In this connection, it is useful to remind ourselves that, away from disk
center, $b_\|$ does begin to contribute more strongly to the determination
of $b_\theta$ and $b_\phi$.
One should therefore calculate the complex polarization not from
$b_\theta$ and $b_\phi$, but from the two components of the field
vector $\bb_\perp$ that is perpendicular to the line of sight.
This would obviously be another next important step to take.
Likewise, it would be highly valuable to inspect the spatial properties
of $E$ and $B$ in much more detail.
This would allow us to study the connection between the sign of $E$
and the topology or structures, and of course between the sign of $B$
and the hemispheric position.

One of the other potential applications of the $EB$ transformation
lies in its potential benefit when regularizing the observed linear
polarization signal.
One could imagine that, instead of applying a random disambiguation for
weak field strengths, one could adopt some type of image reconstruction
in $EB$ space instead of working in $QU$ or $\bb$ space.
This has not yet been explored and would be a useful target for future
research.

Finally, one may wonder whether the global two-scale helicity proxy
introduced here can be used beyond solar physics.
The answer is probably yes, if one thinks about the technique
of Zeeman Doppler imaging of stellar magnetic fields \cite[see,
e.g.,][]{Donati97,Carroll12,Rosen15}.
Likewise, the magnetic field of our own Galaxy may also be subject to such
an analysis \citep{JF12}.
We therefore expect that these points provide exciting opportunities
for future work.

\acknowledgments
This work was carried out in large parts at the Aspen Center for Physics,
which is supported by National Science Foundation grant PHY-1607611.
I thank Evan Scannapieco for organizing the Aspen program on
the Turbulent Life of Cosmic Baryons.
I also thank Gherardo Valori and Etienne Pariat for organizing the Magnetic
Helicity in Astrophysical Plasmas Team at the International Space Science
Institute in Bern, where some early elements of this work were conceived,
and I also thank Maarit K\"apyl\"a, Alexei Pevtsov, Ilpo Virtanen,
and Nobumitsu Yokoi for providing a splendid atmosphere at the
Nordita program on Solar Helicities in Theory and Observations.
I am grateful to Patrik Sanila for help with the spin-weighted
spherical harmonics, and to Ameya Prabhu for showing me some of his
preliminary results with $\pi$-ambiguated magnetic fields using local
patches around specific active regions.
I also thank Marc Kamionkowski and Kandaswamy Subramanian for useful
discussions on the subject, and an anonymous referee for suggesting
improvements to the paper.
This research was supported in part by the Astronomy and Astrophysics
Grants Program of the National Science Foundation (grant 1615100),
and the University of Colorado through
its support of the George Ellery Hale visiting faculty appointment.
I acknowledge the allocation of computing resources provided by the
Swedish National Allocations Committee at the Center for Parallel
Computers at the Royal Institute of Technology in Stockholm.



\begin{thebibliography}{}

\bibitem[Alissandrakis \& Chiuderi-Drago(1994)]{ACD94}
Alissandrakis, C. E., \& Chiuderi-Drago, F.\yapj{1994}{428}{L73}

\bibitem[Blackman \& Brandenburg(2003)]{BB03}
Blackman, E. G., \& Brandenburg, A.\yapjl{2003}{584}{L99}

\bibitem[Bourdin et al.(2018)]{BSB18}
Bourdin, Ph.-A., Singh, N. K., \& Brandenburg, A.\yapj{2018}{869}{3}

\bibitem[Bracco et al.(2019)]{Bracco19}
Bracco, A., Candelaresi, S., Del Sordo, F., \& Brandenburg, A.\yana{2019}{621}{A97}

\bibitem[Brandenburg(2019)]{Bra19AD}
Brandenburg, A.\ymn{2019}{487}{2673}

\bibitem[Brandenburg \& Stepanov(2014)]{BS14}
Brandenburg, A., \& Stepanov, R.\yapj{2014}{786}{91}

\bibitem[Brandenburg et al.(2017a)]{BAJ17}
Brandenburg, A., Ashurova, M. B., \& Jabbari, S.\yapjl{2017a}{845}{L15}

\bibitem[Brandenburg et al.(2019)]{BBKMR19}
Brandenburg, A., Bracco, A., Kahniashvili, T., Mandal, S., Roper Pol, A., Petrie, G. J. D., \& Singh, N. K.\yapj{2019}{870}{87}
(BBKMRPS)

\bibitem[Brandenburg et al.(2009)]{BCC09}
Brandenburg, A., Candelaresi, S., \& Chatterjee, P.\ymn{2009}{398}{1414}

\bibitem[Brandenburg et al.(2002)]{BDS02}
Brandenburg, A., Dobler, W., \& Subramanian, K.\yan{2002}{323}{99}

\bibitem[Brandenburg et al.(2017b)]{BPS17}
Brandenburg, A., Petrie, G. J. D., \& Singh, N. K.\yapj{2017b}{836}{21}
(BPS)

\bibitem[Brandenburg et al.(2011)]{BSBG11}
Brandenburg, A., Subramanian, K., Balogh, A., \& Goldstein, M. L.\yapj{2011}{734}{9}

\bibitem[Carroll et al.(2012)]{Carroll12}
Carroll, T. A., Strassmeier, K. G., Rice, J. B., \& K\"unstler, A.\yana{2012}{548}{A95}

\bibitem[Donati et al.(1997)]{Donati97}
Donati, J.-F., Semel, M., Carter, B. D., Rees, D. E., \& Collier Cameron, A.\ymn{1997}{291}{658}

\bibitem[Durrer(2008)]{Dur08}
Durrer, R.\ybook{2008}{The Cosmic Microwave Background, Chapter 5}
{Cambridge University Press, Cambridge, United Kingdom, 2008}

\bibitem[Jansson \& Farrar(2012)]{JF12}
Jansson, R., \& Farrar, G. R.\yapj{2012}{757}{14}

\bibitem[Georgoulis(2005)]{Geo05}
Georgoulis, M. K.\yapj{2005}{629}{L69}

\bibitem[Goldberg et al.(1967)]{Goldberg67}
Goldberg, J. N., Macfarlane, A. J., Newman, E. T., Rohrlich, F., \& Sudarshan, E. C. G.\yjmp{1967}{8}{2155}

\bibitem[Hoeksema et al.(2014)]{Hoe14}
Hoeksema, J. T., Liu, Y., Hayashi, K., Sun, X., Schou, J., Couvidat, S., Norton, A., Bobra, M., Centeno, R., Leka, K. D., Barnes, G., \& Turmon, M.\ysph{2014}{289}{3483}

\bibitem[Horellou \& Fletcher(2014)]{HF14}
Horellou, C., \& Fletcher, A.\ymn{2014}{441}{2049}

\bibitem[Hughes et al.(2016)]{Hughes}
Hughes, A. L. H., Bertello, L., Marble, A. R., Oien, N. A., Petrie, G., \& Pevtsov, A. A.\arxiv{2016}{1605.03500}

\bibitem[Kahniashvili \& Ratra(2005)]{KR05}
Kahniashvili, T., \& Ratra, B.\yprd{2005}{71}{103006}

\bibitem[Kahniashvili et al.(2014)]{KMLK14}
Kahniashvili, T., Maravin, Y., Lavrelashvili, G., \& Kosowsky, A.\yprd{2014}{90}{083004}

\bibitem[Kamionkowski et al.(1997)]{Kamion97}
Kamionkowski, M., Kosowsky, A., \& Stebbins, A.\yprl{1997}{78}{2058}

\bibitem[Kamionkowski \& Kovetz(2016)]{KK16}
Kamionkowski, M., \& Kovetz, E. D.\yaraa{2016}{54}{227}

\bibitem[Kleeorin et al.(2003)]{KKMRSZ03}
Kleeorin, N., Kuzanyan, K., Moss, D., Rogachevskii, I., Sokoloff, D., \& Zhang, H.\yana{2003}{409}{1097}

\bibitem[Krause \& R\"adler(1980)]{KR80}
Krause, F., \& R\"adler, K.-H.\ybook{1980}
{Mean-field Magneto\-hydro\-dy\-na\-mics and Dynamo Theory}{Oxford: Pergamon Press}

\bibitem[Liu et al.(2017)]{Liu17}
Liu, Y., Hoeksema, J. T., Sun, X., \& Hayashi, K.\ysph{2017}{292}{29}

\bibitem[Pevtsov et al.(1995)]{PCM95}
Pevtsov, A. A., Canfield, R. C., \& Metcalf, T. R.\yapj{1995}{440}{L109}

\bibitem[Pipin \& Pevtsov(2014)]{PP14}
Pipin, V. V., \& Pevtsov, A. A.\yapj{2014}{789}{21}

\bibitem[Pipin et al.(2019)]{PPLK19}
Pipin, V. V., Pevtsov, A. A., Liu, Y., \& Kosovichev, A. G.\yapjl{2019}{877}{L36}

\bibitem[Planck Collaboration Int.\ XXX(2016)]{Adam16}
Planck Collaboration Int.\ XXX.\yana{2016}{586}{A133}

\bibitem[Planck Collaboration results XI(2018)]{Akrami18}
Planck Collaboration results XI.\dana{2018}{10.1051/0004-6361/201832618}{1801.04945}

\bibitem[Roberts \& Soward(1975)]{RS75}
Roberts, P. H., \& Soward, A. M.\yan{1975}{296}{49}

\bibitem[Ros\'en et al.(2015)]{Rosen15}
Ros\'en, L., Kochukhov, O., \& Wade, G. A.\yapj{2015}{805}{169}

\bibitem[Rudenko \& Anfinogentov(2014)]{RA14}
Rudenko, G.~V., \& Anfinogentov, S.~A.\ysph{2014}{289}{1499}

\bibitem[Sakurai et al.(1985)]{Sak85}
Sakurai, T., Makita, M., \& Shibasaki, K.\yproc{1985}{313}
{Theoretical Problems in High Resolution Solar Physics, Proceedings of the MPA/LPARL Workshop held 16-18 September 1985 in M\"unchen, Germany}{H.U Schmidt}{Garching: Max-Planck Institut für Physik und Astrophysik}

\bibitem[Scannapieco \& Ferreira(1997)]{SF97}
Scannapieco, E. S., \& Ferreira, P. G.\yprd{1997}{56}{R7493}

\bibitem[Sc\'occola et al.(2004)]{SHM04}
Sc\'occola, C., Harari, D., \& Mollerach, S.\yprd{2004}{70}{063003}

\bibitem[Seehafer(1990)]{See90}
Seehafer, N.\ysph{1990}{125}{219}

\bibitem[Seljak \& Zaldarriaga(1997)]{SZ97}
Seljak, U., \& Zaldarriaga, M.\yprl{1997}{78}{2054}

\bibitem[Singh et al.(2018)]{Singh18}
Singh, N. K., K\"apyl\"a, M. J., Brandenburg, A., K\"apyl\"a, P. J., Lagg, A., \& Virtanen, I.\yapj{2018}{863}{182}

\bibitem[Sokoloff et al.(1998)]{Soko98}
Sokoloff, D. D., Bykov, A. A., Shukurov, A., Berkhuijsen, E. M., Beck, R., \& Poezd, A. D.\ymn{1998}{299}{189}

\bibitem[Volegova \& Stepanov(2010)]{VS10}
Volegova, A. A., \& Stepanov, R. A.\yjetp{2010}{90}{637}

\bibitem[Warnecke et al.(2011)]{WBM11}
Warnecke, J., Brandenburg, A., \& Mitra, D.\yana{2011}{534}{A11}

\bibitem[Warnecke et al.(2012)]{WBM12}
Warnecke, J., Brandenburg, A., \& Mitra, D.\yjour{2012}{J. Spa.\ Weather Spa.\ Clim.}{2}{A11}

\bibitem[Zaldarriaga \& Seljak(1997)]{ZS97}
Zaldarriaga, M. \& Seljak, U.\yprd{1997}{55}{1830}

\bibitem[Zhang et al.(2010)]{ZSPGXSK10}
Zhang, H., Sakurai, T., Pevtsov, A., Gao, Y., Xu, H., Sokoloff, D. D., \& Kuzanyan, K.\ymn{2010}{402}{L30}

\end{thebibliography}
\end{document}